\documentclass[aps, prl, nofootinbib, twocolumn, superscriptaddress, noeprint, nolongbibliography]{revtex4-2}

\emergencystretch=\maxdimen
\usepackage[mathlines]{lineno}

\usepackage{amsmath}
\usepackage{amssymb}
\usepackage{bm}         
\usepackage{braket}
\usepackage{dcolumn}    
\usepackage{diagbox}
\usepackage{epsfig}
\usepackage{hhline}
\usepackage{mathrsfs}
\usepackage{newtxmath, newtxtext}
\usepackage{slashed}
\usepackage[dvipsnames]{xcolor}
\usepackage{mathtools, cuted}
\usepackage{lipsum}
\usepackage{tabularx}
\usepackage{tabularray}
\usepackage{nicematrix}
\usepackage{wrapfig,float}
\usepackage{multirow}
\usepackage{hhline}
\usepackage{colortbl}
\usepackage{comment}
\usepackage{graphicx}
\usepackage{subcaption}
\usepackage{ragged2e}
\usepackage{setspace}

\usepackage[pagebackref=false, colorlinks=true]{hyperref}
\definecolor{reddish}{rgb}{0.7,0.2,0.0}
\definecolor{blueish}{rgb}{0.1,0.1,1}
\hypersetup{
linkcolor=reddish,      
citecolor=blueish,      
filecolor=blue,         
urlcolor=magenta}       

\newcommand{\rp}{r_\text{p}}
\renewcommand{\Pr}{P_{\!(r)}} 
\newcommand{\Pphi}{P_{\!(\varphi)}}
\newcommand{\Ptheta}{P_{\!(\vartheta)}}
\renewcommand{\Pi}{P_{\!(i)}} 

\begin{document}

\title{Universal Bounds on Black Hole Observables Imposed by Energy Conditions}

\author{Rahul Kumar Walia}\email{rahul.phy3@gmail.com}
\affiliation{Mullard Space Science Laboratory, University College London, Holmbury St. Mary, Dorking, Surrey RH5 6NT, UK}
\affiliation{Department of Physics, Netaji Subhash University of Technology, New Delhi 110078, India}
\author{Boris Georgiev}
\affiliation{Steward Observatory and Department of Astronomy, University of Arizona, 933 N. Cherry Avenue, Tucson, AZ 85721, USA}
\author{Chi-kwan Chan}
\affiliation{Steward Observatory and Department of Astronomy, University of Arizona, 933 N. Cherry Avenue, Tucson, AZ 85721, USA}
\affiliation{Data Science Institute, University of Arizona, 1230 N. Cherry Avenue, Tucson, AZ 85721, USA}
\affiliation{Program in Applied Mathematics, University of Arizona, 617 North Santa Rita, Tucson, AZ 85721, USA}

\begin{abstract}
Energy conditions (ECs) underpin fundamental results in general relativity, yet their observational validity in strong gravitational fields remains largely untested. Here, we establish a direct correspondence between the local stress-energy tensor, photon-orbit dynamics, and strong-field observables associated with black hole shadows, photon rings, and eikonal gravitational-wave ringdown for static, spherically symmetric black holes within general relativity. We show that imposing local ECs at unstable photon spheres yields analytic bounds on these observables. In particular, the demagnification of successive photon rings provides an exact diagnostic of the local tangential null energy condition. These bounds offer a direct route to constraining the matter fields sourcing spacetimes and testing ECs through strong-field observations. Combining them with observational measurements of these observables further restricts admissible black hole geometries and additionally constrains the otherwise inaccessible photon sphere geometry and local stress-energy content, demonstrating their observational potential in the strong-field regime.
\end{abstract}

\keywords{General Relativity---Non-Kerr Spacetimes}
\maketitle

\emph{Introduction---}%
Einstein's field equations (EFEs) relate spacetime geometry to its matter-energy content. However, without imposing restrictions on the latter, they admit solutions exhibiting pathological features such as closed timelike curves, naked singularities, and traversable wormholes~\cite{Misner:1973prb}. Classical energy conditions (ECs) impose physically motivated restrictions on gravitating matter fields and thus play a key role in excluding \textit{unphysical} spacetimes in general relativity (GR)~\cite{Curiel:2014zba,Kontou:2020bta}.

ECs led to several seminal results in a variety of contexts within GR, including gravitational collapse, black hole (BH) formation, dynamical stability of spacetimes, causal structure, and spacetime singularity theorems~\citep{1955PhRv...98.1123R,Hawking:1973uf,Penrose:1969pc}. Beyond these theoretical applications, ECs have also been widely used to derive model-independent bounds on cosmological observables, enabling direct tests against large-scale observations and constraining both cosmic matter content and modified theories of gravity~\cite{Visser:1997qk,Lima:2008sq}. Yet, their direct empirical tests in astrophysical strong-field regimes remains largely unexplored, despite their fundamental role in BH physics and the profound implications that any violation would have for our understanding of gravity. 

Observables associated with photon orbits offer a promising avenue for such tests. BHs possess photon spheres (PSs), formed by constant-radius unstable photon orbits~\cite{Chandrasekhar:1985kt}, which govern strong-field observable signatures such as BH shadows, photon rings, and gravitational waves (GWs) ringdown in the eikonal limit~\cite{Cardoso:2008bp,EventHorizonTelescope:2019dse,2020SciA....6.1310J}. More generally, non-vacuum BHs and certain horizonless ultracompact objects (HUCOs) can support multiple PSs, including stable ones~\cite{Shipley:2023umh,Cunha:2020azh,Dolan:2016bxj,Cardoso:2014sna,Guo:2022ghl}.

Despite the advent of multi-messenger observations probing astrophysical BHs, from stellar to supermassive scales \cite{LIGOScientific:2016aoc,EventHorizonTelescope:2019dse}, ECs remain largely untested in strong gravitational fields, where their validity is typically assumed rather than empirically assessed. However, testing ECs requires quantitative observational predictions that can be directly compared with observations, which are currently lacking. In this Letter, we bridge precisely this gap. We develop a general framework that translates local strong-field ECs into quantitative constraints on observables in static, spherically symmetric spacetimes within GR, thereby opening a pathway to direct observational tests of ECs.

We first show that the existence, multiplicity, and (in)stability of PSs in BH and HUCO spacetimes can be expressed directly in terms of local energy density and pressures of the background fields sourcing these spacetimes. Imposing the ECs further constrains the compactness and stability of PSs. Then, by demanding that these gravitating fields satisfy all ECs at unstable PSs (UPSs), we derive analytic, model-independent bounds on several key observables, including the shadow size, photon ring observables such as the Lyapunov exponent and time delay, and the eikonal QNM frequencies through the Lyapunov timescale. The derived bounds are immediately applicable to current and forthcoming observations by the Event Horizon Telescope (EHT)~\cite{EventHorizonTelescope:2019dse}, future space-based Very Long Baseline Interferometry (VLBI) missions, Black Hole Explorer (BHEX)~\cite{Johnson:2024ttr}, and GW observations by the Laser Interferometer Space Antenna (LISA)~\cite{Barausse:2020rsu}. This establishes PS phenomenology as a direct observational probe of the ECs.

Furthermore, comparing these theoretical bounds with observations offers a unified framework for simultaneously constraining the entire class of spherically symmetric spacetimes, their underlying matter sources, and the validity of the ECs. This also overcomes the limitations of existing approaches to constraining BH spacetimes, which generally rely on metric-by-metric analyses or numerical parameter estimation, rendering their constraints model-dependent and often subject to degeneracies between gravitational and astrophysical effects. We use natural units with $c = G = 1$.


\emph{Spherically Symmetric Static Spacetime---}
We consider static and spherically symmetric parameterized spacetime, which is expressed in areal-radius coordinates $(t,r,\vartheta,\varphi)$ as follows:
\begin{align}
{\rm d}s^2= -A(r) {\rm d}t^2 +B(r){\rm d}r^2 +r^2 ({\rm d}\vartheta^2 + \sin^2\vartheta\,{\rm d}\varphi^2).\label{Eq:Static-Metric}
\end{align}
We impose the following key assumptions: \textit{i}) the analysis is restricted to the static spacetime region where the metric remains Lorentzian and non-singular, i.e., $A, B>0$;
\textit{ii}) the metric~\eqref{Eq:Static-Metric} represents both BH and HUCO solutions of the EFEs; 
\textit{iii}) the photon field is minimally coupled to gravity and other background fields, ensuring that photons follow null geodesics of the metric~\eqref{Eq:Static-Metric}, and \textit{iv}) gravitating fields are assumed not to dynamically back-react, so photons propagate on a fixed background spacetime.
Such photons with impact parameter $\eta$ follow radial equation of motion $A B\, \dot{r}^2+\mathcal{V}=1$. In asymptotically flat BH spacetimes, the radial potential $\mathcal{V}=A \eta^2/r^2$, vanishes at both the horizon and spatial infinity, with positive values in between, guaranteeing at least one extremum. When a single extremum exists, it must be a global \emph{maximum}, corresponding to an UPS with $\mathcal{V}=1$, $\mathcal{V}'=0$, and $\mathcal{V}''<0$, where photons move on a circular orbit of constant-radius, $\rp$, with $\dot{r}=\ddot{r}=0$. Here, dot and prime, respectively, represent derivatives with respect to the affine parameter $\tau$ and $r$. When multiple extrema exist, additional PSs always appear in pairs of one stable and one unstable~\cite{Cunha:2017qtt}
\footnote{For HUCO, the numbers of SPS and UPS depend on the nature of central region of HUCO. 
}.

It follows directly from metric~\eqref{Eq:Static-Metric} that the PS radius $\rp$, its stability under radial perturbations, and the shadow radius $\eta_{\rm p}$---corresponding to the impact parameter of ingoing photons asymptotically approaching an UPS%
\footnote{In the presence of multiple PSs, the shadow radius is determined by the smallest impact parameter $\eta_{\rm p}$ through which photons approach one of the UPS \cite{Guo:2022ghl}.}%
---are determined by:
\begin{equation}\label{Eq:PS-Conditions}
\begin{aligned}
 \text{PS Radius~$\rp$:}&\quad \mathscr{f}(\rp) \equiv \left.\left(\frac{r^2}{A}\right)'\right|_{\rp}=0,\\
 \text{PS (in)Stability:}&\quad \mathscr{g}(\rp) \equiv \left.\left(\frac{A}{r}\right)^2\left(\frac{r^2}{A}\right)''\right|_{\rp} (>) < 0,\\
 \text{Shadow Radius:}&\quad \eta_{\rm p} = \left.\sqrt{\frac{r^2}{A}}\right|_{\rp}.
\end{aligned}
\end{equation}
Henceforth, quantities with subscript ``p'' are evaluated at PS.

Photons closely following the UPS produce multiple images labeled by the number $n$ of half-orbits around the ultracompact object. For an extended source, these images form the \textit{photon ring}, a sequence of increasingly demagnified and time-delayed subrings that converge exponentially to the shadow boundary $\eta_{\rm p}$~\cite{2020SciA....6.1310J}. Higher-order subrings follow universal scaling relations in terms of lensing observables~\cite{2020SciA....6.1310J, Kocherlakota:2024hyq}: $F_{n+2}/F_n\approx e^{-2\gamma_{\rm p}}$, $t_{n+2}-t_n \approx 2t_{\rm d;p}$. Here, the Lyapunov exponent $\gamma_{\rm p}=\pi \eta_{\rm p}\sqrt{\mathscr{g}_{\rm p}/2 B_{\rm p}}$ and delay time $t_{\rm d;p}=\pi \eta_{\rm p}$, respectively, control the demagnification or exponential suppression of subring flux $F_n$ and the elapsed time in appearance of $n^{\rm th}$ and $(n+2)^{\rm th}$ subrings~\cite{2020SciA....6.1310J, Kocherlakota:2024hyq}. While $\gamma_{\rm p}$ quantifies the radial instability of the PS, i.e., how quickly light leaks out of the UPS, the instability timescale is set by the Lyapunov time $t_{\rm \ell;p}=\sqrt{2B_{\rm p}/\mathscr{g}_{\rm p}}$ \cite{Kocherlakota:2024hyq}.
Additionally, $\eta_{\rm p}$ and $t_{\rm \ell;p}$ govern QNM frequencies in the eikonal limit $\omega_{l \bar{n}}=l/\eta_{\rm p} -i\left(\bar{n}+1/2\right)/t_{\rm \ell;p}$~\cite{Cardoso:2008bp},
where $\bar{n}$ and $l$ are the overtone and angular momentum numbers. Since $t_{\rm \ell;p} \gamma_{\rm p} = \pi\eta_{\rm p}$, any two of these observables determine the third. Note that a joint measurement of the shadow size and at least one photon-ring observable can completely break degeneracies in BH parameter estimation.

Note that the observables, $\eta_{\rm p}$, $\gamma_{\rm p}$, $t_{\rm d;p}$, and $t_{\ell;\rm p}$, are determined by the metric functions at the UPS. We use ECs to constrain the spacetime geometry at the UPS and thus, in turn, to establish stringent bounds on the BH observables. Moreover, these geometric observables are largely insensitive to astrophysical uncertainties and are accessible to current and forthcoming VLBI and GW observations, therefore, they provide clean probes of the near-UPS geometry~\cite{KumarWalia:2024omf}. While current EHT measurements of the shadow size primarily constrain the metric function $A$~\cite{EventHorizonTelescope:2020qrl}, future photon-ring observations will additionally probe $B$, providing a direct observational test of the EC-driven constraints on the near-UPS spacetime geometry and the observables derived here.

\emph{EC-Driven Bounds On Observables---}
In the local inertial frame, defined by the orthonormal tetrad $\mathbf{e}^{\mu}_{(a)}=|g_{aa}|^{-1/2}\delta^{\mu}_a$, the stress-energy tensor of the matter fields is $\mathcal{G}_{(a)(b)}
= \mathcal{G}_{\mu\nu}\mathbf{e}^{\mu}_{(a)}\mathbf{e}^{\mu}_{(b)}
= 8\pi\,\mathcal{T}_{(a)(b)}
= 8\pi\,\text{diag} \{\rho, \Pr, \Ptheta, \Pphi\}$,
where $\rho$ is the locally measured energy density and $\Pi$ are the principal pressures:
\begin{equation}\label{Eq:EMT}
\begin{aligned}
    \rho &= \frac{1}{8\pi r^2}\bigg[1-\left(\frac{r}{B}\right)'\bigg],\quad
    \Pr = \frac{1}{8\pi r^2}\bigg[\frac{(rA)'}{AB}-1\bigg],\\
    \Ptheta &= \Pphi = \frac{1}{32\pi r}\bigg[\left(\frac{rA'}{AB}\right)'+ \frac{(rA')'}{AB}+2\left(\frac{1}{B}\right)'\bigg].
\end{aligned}
\end{equation}
It is useful to define the Misner-Sharp-Hernandez (MSH) mass $m(r)$, a quasi-local measure of gravitational energy~\cite{1964PhRv..136..571M,1966ApJ...143..452H}, via $B=(1 - 2m/r)^{-1}$, such that $m(r)$ matches the Arnowitt-Deser-Misner (ADM) mass, $M$, at spatial infinity. Positive energy density, $\rho=m'/4\pi r^2 >0$, ensures that $m(r) \leq M$.

The existence and radial (in)stability of a PS are encoded entirely in $\mathcal{T}_{(a)(b)}$: 
\begin{eqnarray}\label{Eq:PS-Conditions1}
     \text{PS Radius~$\rp$:}&&~~\!\!\left.\Pr\right|_{\rp} = \left.\frac{1}{4\pi r^2}\left(1-\frac{3 r_{\rm h}}{2r}-\frac{3}{r}\int_{r_{\rm h}}^{r}4\pi \Tilde{r}^2\rho(\Tilde{r})d\Tilde{r}\right)\right|_{\rp}, \nonumber\\
    \text{PS (in)Stability:}&&~~\!\!\left.\left(\rho + \Ptheta\right)\right|_{\rp} (<)> \frac{1}{8\pi\rp^2}.
\end{eqnarray}
The event horizon is determined by solving $r_{\rm h}=2m(r_{\rm h})$, where $1/B\to0$ and $\left.(\rho + \Pr)\right|_{r_{\rm h}} \rightarrow 0$, provided that $A'$ and $B'$ remain finite at the horizon~\cite{Hod:2013jhd}. Equation~(\ref{Eq:PS-Conditions1}) provides deep insights into strong-field gravity: A local matter-driven mechanism governing PS formation and the transition between SPS and UPS. In particular, the PS radius depends only on $\rho$ and $P_{(r)}$, independently of the tangential pressures, whereas its (in–)stability is controlled locally and exclusively by $\rho$ and $P_{(\vartheta)}$; see Ref.~\cite{mycomment}. An immediate consequence is that spacetimes that violate or saturate the tangential Null EC, $(\rho + \Ptheta)\leq 0$, cannot support an SPS. This explains the absence of SPS in vacuum BH spacetimes. Moreover, since $\rho$ is independent of the metric function $A$ [see Eq.~(\ref{Eq:EMT}], PSs may also exist in spacetimes with locally negative energy density, as in phantom field solutions~\cite{Eiroa:2013nra}. Equation~(\ref{Eq:PS-Conditions1}) therefore provide a model-independent characterization of PS directly in terms of the matter content, without requiring the explicit metric. 

Similarly, BH observables can be determined directly from the given stress-energy tensor components:
\begin{equation}\label{Eq:BH_Obs}
\begin{aligned}
     \eta_{\rm p} \!&=\! \sqrt{\frac{3r_{\rm p}^2}{1+8\pi r_{\rm p}^2\,P_{(r);\rm p}}}\exp\left[\int_{\rp}^{\infty}\!\!\frac{ 4\pi r^2(\rho+P_{(r)})}{r-8\pi\int_{0}^{r} \Tilde{r}^2\rho\, {\rm d}\Tilde{r}}{\rm d}r\right],\\
    \gamma_{\rm p} \!&=\! \pi\sqrt{1-8\pi r_{\rm p}^2\, (\rho_{\rm p} + P_{(\vartheta);\rm p})}\,,\\
    t_{\ell; \rm p} \!&=\! \frac{\eta_{\rm p}}{\sqrt{1-8\pi r_{\rm p}^2 (\rho_{\rm p} + P_{(\vartheta);\rm p})}},
\end{aligned}    
\end{equation}
and $t_{\rm d;p}=\pi\, \eta_{\rm p}$; note that $P_{(\vartheta)}=P_{(\varphi)}$. In vacuum, these observables reduce to their Schwarzschild values: $\eta_{\rm p}=3\sqrt{3}M,\, \gamma_{\rm p}=\pi, \, t_{\rm d;p}=3\sqrt{3}\pi M$, $t_{\rm \ell;p}=3\sqrt{3} M$ \cite{Cardoso:2014sna,Kocherlakota:2024hyq}. Crucially, different observables probe distinct combinations of the stress-energy components: while $\eta_{\rm p}$ depends on the local radial pressure and the integrated exterior matter distribution, $\gamma_{\rm p}$ is determined solely by the tangential NEC at the UPS. This suggests a route toward distinguishing gravitating matter fields around compact objects, identifying observational degeneracies, and testing local ECs using strong-field observables. For instance, BHs with matter fields satisfying $\rho+P_{(\vartheta)}=0$ (e.g., static scalar field models) are exactly degenerate with Schwarzschild BH in photon–ring demagnification, $\gamma_{\rm p}=\pi=\gamma^{\rm Schw}_{\rm p}$. Thus, a measurement of $\gamma_{\rm p}=\pi$ does not uniquely imply a Schwarzschild BH. However, the shadow size, subrings time delay, and GW ringdown may still break this degeneracy. Eq.~(\ref{Eq:BH_Obs}) infers that a comparatively modest shadow size can coexist with an anomalously short/long instability timescale.

ECs imposed on local matter fields at UPS further restrict the allowed values of these observables. We consider four pointwise ECs \cite{Curiel:2014zba}%
\footnote{Several lesser-known ECs---typically formulated as variations of the standard ones discussed above---impose additional constraints on the stress-energy tensor~\cite{Kontou:2020bta,Curiel:2014zba}.}:
\textit{i})~the Null EC (NEC): $\rho + \Pi \geq 0$;
\textit{ii})~the Weak EC (WEC): $\rho \geq 0$ and $\rho + \Pi \geq 0$;
\textit{iii})~the Strong EC (SEC): $\rho + \Pi \geq 0$ and $\rho + \sum_i \Pi \geq 0$; and
\textit{iv})~the Dominant EC (DEC): $\rho \geq 0$ and $\rho \geq |\Pi|$. These ECs lead to several interesting consequences.

An intriguing result follows from Eq.~(\ref{Eq:BH_Obs}). BHs sourced by matter fields satisfying the WEC at the UPS (e.g., electromagnetic fields and potential-dominated canonical scalar fields) have $-\rho_{\rm p} \leq P_{(r);\rm p} \leq 0$ and $0\leq 8\pi\rp^2\left(\rho_{\rm p} + P_{(\vartheta);\rm p}\right) < 1$, and as a result, $\eta_{\rm p}\geq \sqrt{3}r_{\rm p}$, $\gamma_{\rm p}\leq \pi$ and $t_{\ell; \rm p}\geq \eta_{\rm p}$, equivalently, $t_{\ell; \rm p}\geq t_{\rm d; p}/\pi$. Remarkably, $\gamma_{\rm p}\leq\pi$ is both necessary and sufficient for the tangential NEC at the UPS, independently of the exterior matter distribution. $\gamma_{\rm p}>\pi$ would indicate a local violation of the tangential NEC at the UPS. The photon ring demagnification and the eikonal QNM damping rate ($\propto 1/{t_{\ell; \rm p}}$) are universally bounded from above by their Schwarzschild values. Thus, a measured violation of these bounds would directly signal WEC violation within the broad class of static, spherically symmetric spacetimes.

Physically, a strictly positive tangential NEC reduces the radial instability of neighboring light rays per orbit ($\gamma_{\rm p}<\pi$), allowing them to remain trapped longer and complete more orbits before escaping. As a result, this reduces the geometric suppression of successive photon subrings, potentially producing wider and brighter subrings than in Schwarzschild case. Equivalently, for a fixed $\eta_{\rm p}$, a smaller $\gamma_{\rm p}$ implies a smaller QNM damping rate and therefore longer-lived ringdown modes, with a larger quality factor than the corresponding Schwarzschild value, e.g., the fundamental quadrupolar mode quality factor $Q_{20}\geq 2$. Schwarzschild and non-vacuum BHs satisfying $\rho + \Ptheta=0$ at the UPS possess maximally unstable PS, i.e., the greatest light leakage per orbit among all NEC-satisfying spacetimes. Photon ring observations in BH images and measurements of the damping-to-oscillation ratio in GW signals can test the tangential NEC in the strong-field regime.

Imposing all four ECs at an UPS (see Tables \ref{Table-Static2} and \ref{Table-Static3}) further yields the bounds
\begin{equation}
    \text{max}\!\left[0, \frac{2 \left(B_{\rm p}-r_{\rm p} B_{\rm p}'\right)}{B_{\rm p}^2}-1 \right] < \frac{\gamma_{\rm p}^2}{\pi^2} \leq \left( \frac{4B_{\rm p}-r_{\rm p}B_{\rm p}'}{2B_{\rm p}^2} \right),\label{Eq:Bound_Lyapunov1}\\
\end{equation}    
and $\text{max}\left(-2, 1-B_{\rm p}+|B_{\rm p}-3|\right)\leq
            r_{\rm p}B_{\rm p}'/B_{\rm p}$, or equivalently, 
\begin{equation}
    \text{max}\!\left[0, 1- 16 \pi r_{\rm p}^2\,\rho_{\rm p} \right]< \! \frac{\gamma_{\rm p}^2}{\pi^2} \!\leq\!\! \left(1 - 4 \pi r_{\rm p}^2(\rho_{\rm p}-P_{(r);\rm p})\right), \label{Eq:Bound_Lyapunov2}
\end{equation}
and $|P_{(r);\rm p}|\leq \rho_{\rm p}$. The radial DEC further ensures that  $\gamma_{\rm p}\leq\pi$. Similarly, for asymptotically flat spacetimes supporting an UPS, the shadow radius satisfies $\sqrt{3}\rp \leq \eta_{\rm p}\leq 3\sqrt{3}M$ \cite{Cvetic:2016bxi, Yang:2019zcn}. Combining them establishes bounds on the time-domain observables, namely the photon ring delay time and the Lyapunov time:
\begin{eqnarray}
        \sqrt{3}\pi\rp \leq &\, t_{\rm d;p}\, & \leq 3\sqrt{3} \pi M,\label{Eq:Bound_Delay_Time1}\\
        \frac{6r_{\rm p}^2 B_{\rm p}^2}{4B_{\rm p}-r_{\rm p}B_{\rm p}'} \leq & t_{\rm \ell;p}^2 & < \! \frac{27M^2}{\text{max}\!\left[0, \frac{2 \left(B_{\rm p}-r_{\rm p} B_{\rm p}'\right)}{B_{\rm p}^2}-1 \right]},\ \ \  \label{Eq:Bound_Lyapunov_Time1}
\end{eqnarray}
and equivalently, in terms of $\rho$ and $P_{(r)}$:
\begin{eqnarray}
    \!\!\!\!\frac{3r_{\rm p}^2}{1-4 \pi r_{\rm p}^2(\rho_{\rm p}-P_{(r);\rm p}) }\!\leq & t_{\rm \ell;p}^2 & < \frac{27 M^2}{\text{max}\left[0, 1 - 16 \pi r_{\rm p}^2\,\rho_{\rm p} \right]},\label{Eq:Bound_Lyapunov_Time2}
\end{eqnarray}
where $r_{\rm p}$ is determined by solving Eq.~\eqref{Eq:PS-Conditions1}. These results demonstrate that ECs leave direct and measurable observational imprints. PS observables therefore become local probes of the matter fields supporting the spacetime, strong-field gravity, and eventually of the ECs. 

These universal bounds on observables—expressed equivalently in terms of either the local spacetime metric functions or the local stress-energy components—hold for all static, spherically symmetric spacetimes in GR. The later formulation provides a direct observational diagnostics of the matter fields sourcing the BH spacetimes. 

We now specialize to a particular case – \textit{critical} radial NEC, $\rho+P_{(r)}=0$, which yields $A(r)B(r)=\mathrm{const}$ and, upon normalizing the time coordinate, $A(r)B(r)=1$. This case encompasses a broad class of BH and HUCO spacetimes\footnote{The horizon radius of such asymptotically flat BHs with ADM mass $M$ is determined by solving $r_{\rm h}=2M-\int_{r_{\rm h}}^{\infty} 8\pi r^2 \rho\, {\rm d}r$.}. The stress-energy components then simplify to $\rho=-P_{(r)}=m'/(4\pi r^2)$ and $P_{(\vartheta)}=P_{(\varphi)}=-m''/(8\pi r)$. This class leads to particularly simple relations among the PS, the MSH mass, and the local matter fields, while tightening the bounds on observable quantities:
\begin{eqnarray}
    {\rm PS~Radius:}               &&   \left.\rho\right|_{\rp}\!\!=\!\!\left.\frac{1}{4\pi r^2}\left(\frac{3m}{r}-1\right)\right|_{\rp}\,\!\!\! \Rightarrow\,\!\! \rp = \left.3 m_{\rm G}\right|_{\rp},\label{Eq:PS-Radius1}\\
    \!\!\!\!\!\!{\rm PS~(in)Stability:}  &&\; \left.P_{(\vartheta)}\right|_{\rp}(<)\!>\left.\frac{3}{8\pi r^2}\left(1-\frac{2m}{r}\right)\right|_{\rp}\!\!\!,\label{Eq:PS-Stability1}\\
    &&\; \left.\left(\rho+P_{(\vartheta)}\right)\right|_{\rp}(<)\!>\frac{1}{8\pi\,\rp^2},\label{Eq:PS-Stability2} 
\end{eqnarray}
where $m_{\rm G}=\left(m - 4\pi r^3\rho/3\right)$ is an \textit{effective PS mass}. Thus, PS formation is governed not by the enclosed MSH mass alone, but by its local density–corrected value, $m_{\rm G}$. The familiar Schwarzschild relation $\rp=3M$ emerges as the vacuum limit of this quasi-local relation $\rp = 3 m_{\rm G;p}$, valid throughout the critical-radial-NEC class. Geometrically, the PS condition becomes $K_{\rm p}=4\pi\bar{\rho}_{\rm G;p}$, where $K_{\rm p}=1/r_{\rm p}^2$ is the intrinsic Gaussian curvature and $\bar{\rho}_{\rm G;p}=3m_{\rm G;p}/(4\pi r_{\rm p}^3)$ is the effective mean density of PS. Its stability is independently controlled by the local tangential stress through Eq.~(\ref{Eq:PS-Stability2}).

Beyond determining PS radius from a given matter-energy distribution in Eq.~(\ref{Eq:PS-Radius1}), Eqs.~(\ref{Eq:PS-Stability1}) and (\ref{Eq:PS-Stability2}) reverse the problem: PS stability further constrains the admissible matter-energy distribution. In the static region, a SPS necessarily requires positive tangential pressure, $P_{(\vartheta);\rm p}>0$, although this alone does not guarantee stability; an UPS, by contrast, admits either sign. Consequently, matter fields with $P_{(\vartheta);\rm p}\leq 0$ (for e.g., massless scalar fields, certain k-essence and phantom models, minimally coupled scalar fields with suitable potentials, and ghost condensates) at the PS cannot independently support an SPS. More fundamentally, Eqs.~(\ref{Eq:PS-Radius1}) and (\ref{Eq:PS-Stability1}) establish a direct correspondence between the local matter equation–of–state, PS compactness $m_{\rm p}/\rp$, and its stability. 

Remarkably, the energy–density distribution is enough to determine both the PS radius through Eq.~(\ref{Eq:PS-Radius1}) and the shadow size $\eta_{\rm p}=\sqrt{3r_{\rm p}^2/(1-8\pi r_{\rm p}^2\,\rho_{\rm p})}$. Conversely, the shadow size and Lyapunov exponent determine the local matter properties:
\begin{equation}
    \rho_{\rm p} = \frac{1}{8\pi r_{\rm p}^2}\left(1-\frac{3 r_{\rm p}^2}{\eta^2_{\rm p}}\right),\;\;
    P_{(\vartheta);\rm p} = \frac{1}{8\pi r_{\rm p}^2}\left(\frac{3 r_{\rm p}^2}{\eta^2_{\rm p}}-\frac{\gamma_{\rm p}^2}{\pi^2}\right).\label{Eq:PS_Energy_Density}
\end{equation}

Imposing the ECs confines a SPS to $2m_{\rm p}< \rp < (12/5)m_{\rm p}$ and requires $1/2<8\pi r_{\rm p}^2\,\rho_{\rm p} < 1$ and $3\left(1-2m_{\rm p}/r_{\rm p}\right) < 8\pi r_{\rm p}^2\, P_{(\vartheta);\rm p} \leq 2\left(3m_{\rm p}/r_{\rm p}-1\right)$. For an UPS, the corresponding constraints are $2m_{\rm p} < r_{\rm p} \leq 3 m_{\rm p}$,  $0 \leq 8\pi r_{\rm p}^2\,\rho_{\rm p} < 1$ and $0 \leq 8\pi r_{\rm p}^2\, P_{(\vartheta); \rm p}  < \text{min} \left(2\left(3m_{\rm p}/r_{\rm p}-1\right),\, 3\left(1-2m_{\rm p}/r_{\rm p}\right)\right)$. See Table~\ref{Table-Static1} for detailed EC-driven constraints. These bounds constrain the matter distributions needed to support UPS and SPS, providing model-independent criteria for constructing physically admissible BH and HUCO spacetimes.

Although the UPS radius is not directly observable, the ECs constrain it through the observables:
\begin{equation}
    \frac{\eta_{\rm p}}{\sqrt{3}}\frac{\gamma_{\rm p}}{\pi} \leq \rp \leq \frac{\eta_{\rm p}}{\sqrt{6}}\sqrt{1+\frac{\gamma^2_{\rm p}}{\pi^2}}\label{Eq:PS_Radius_Constraint}.
\end{equation}
For this class, Eqs.~(\ref{Eq:Bound_Lyapunov1}) and (\ref{Eq:Bound_Lyapunov_Time1}) reduce to
\begin{eqnarray}
    \text{max}\left[0, 5- \frac{12m_{\rm p}}{r_{\rm p}}\right] < \,\frac{\gamma_{\rm p}^2}{\pi^2} &\, \leq& 3\left(1-\frac{2m_{\rm p}}{r_{\rm p}} \right),\label{Eq:Bound_Lyapunov3} \\
    \frac{r_{\rm p}^2}{1-\frac{2 m_{\rm p}}{r_{\rm p}}} \leq \, t_{\rm \ell;p}^2 & <& \frac{27M^2}{\text{max}\left[0, 5- \frac{12 m_{\rm p}}{r_{\rm p}}\right]},
    \ \ \ 
    \label{Eq:Bound_Lyapunov_time3}
\end{eqnarray}
or, equivalently
\begin{eqnarray}
\max\!\left[0, -1+\frac{6 r_{\rm p}^2}{\eta_{\rm p}^{2}}\right]
< \frac{\gamma_{\rm p}^{2}}{\pi^{2}} & \leq &  \frac{3 r_{\rm p}^2}{\eta_{\rm p}^{2}},\\
\max\!\left[0, \frac{3-8\pi\eta_{\rm p}^{2}\,\rho_{\rm p}}{3+8\pi\eta_{\rm p}^{2}\,\rho_{\rm p}}\right]
< \frac{\gamma_{\rm p}^{2}}{\pi^{2}} & \leq &  \frac{3}{3+8\pi\eta_{\rm p}^{2}\rho_{\rm p}},  \label{Eq:Bound_Lyapunov4}  \\
\eta_{\rm p}^{2}  \leq  t_{\ell;{\rm p}}^{2}  & < &  \frac{27M^{2}} {\max\!\left[0, \frac{3-8\pi\eta_{\rm p}^{2}\,\rho_{\rm p}}{3+8\pi\eta_{\rm p}^{2}\,\rho_{\rm p}} \right]}. \label{Eq:Bound_Lyapunov_time4}
\end{eqnarray}

A particularly restrictive result emerges when the matter fields have nonnegative trace at the UPS. Together with the DEC, this forces $\rho=-P_{(r)}=P_{(\vartheta)}=P_{(\varphi)}$, so that the stress-energy tensor is locally tracefree (for e.g., electromagnetic fields). In this case, the PS radius and local energy density can then be reconstructed entirely from two observables:
\begin{equation}
    \rp=\frac{\eta_{\rm p}}{\sqrt{6}}\sqrt{1+\frac{\gamma_{\rm p}^2}{\pi^2}}, \quad \rho_{\rm p}=\frac{3}{8\pi \eta^2_{\rm p}}\left(\frac{\pi^2-\gamma_{\rm p}^2}{\pi^2+\gamma_{\rm p}^2}\right).
\end{equation}
Admissible UPS is further restricted to $(12/5)m_{\rm p}< \rp \leq 3m_{\rm p}$ and $0 \leq 8\pi r_{\rm p}^2\,\rho_{\rm p} < 1/2$, with shadow size constrained as $\sqrt{3}\rp \leq \eta_{\rm p} < \sqrt{6}\rp$  or $3\sqrt{3} \leq \eta_{\rm p}/m_{\rm p} < 12\sqrt{6}/5$. Thus, joint measurements of the shadow size and photon-ring demagnification can determine the otherwise inaccessible UPS and local matter content. Together, these results establish a direct connection between local matter properties and BH observables, providing model-independent consistency tests of ECs in the strong-field regime.

\emph{Discussion---}
ECs have long been central to GR, yet their validity in astrophysical strong-field regimes remains largely untested. Here, we address this gap by deriving model-independent, EC-driven bounds on strong-field observables of static, spherically symmetric BHs and HUCOs within GR. We first establish general relations connecting PS existence, stability, and key observables associated with UPSs, including the shadow size, Lyapunov exponent (controlling photon-ring demagnification), photon-ring time delay, and Lyapunov time (controlling eikonal GW ringdown damping), to the matter content sourcing the spacetime. These relations reveal that the PS radius and shadow size can be completely determined from the energy density distribution and radial pressure, independently of the tangential pressures, whereas the local values of energy density and tangential pressure determine PS stability and the Lyapunov exponent. Interestingly, the tangential NEC directly controls the instability of photon orbits near UPS. This establishes a direct, model-independent correspondence between matter content and PS dynamics. Imposing ECs at the UPS, in turn, yields analytic and testable bounds on UPS compactness and observables. Confronting these EC-driven bounds with observations offers a pathway to constraining the UPS and local matter fields, and testing ECs in the strong-field regime. 

The shadow size and Lyapunov exponent particularly encode complementary information about the local energy density and tangential pressure. In particular, the Lyapunov exponent provides an exact local diagnostic of the tangential NEC: $\gamma_{\rm p}\leq\pi$ if and only if $\rho_{\rm p}+P_{\vartheta,\rm p}\geq0$. Thus, $\gamma_{\rm p}>\pi$ would directly signal its violation in static, spherically symmetric spacetimes within GR. For the critical-radial-NEC class, ECs further constrain the PS radius directly in terms of the shadow size and Lyapunov exponent. 

More generally, these EC-driven bounds admit equivalent formulations in terms of the metric functions (Eqs.~(\ref{Eq:Bound_Lyapunov1}), (\ref{Eq:Bound_Delay_Time1}), and (\ref{Eq:Bound_Lyapunov_Time1})) and the local stress-energy components of the gravitating fields (Eqs.~(\ref{Eq:Bound_Lyapunov2}) and (\ref{Eq:Bound_Lyapunov_Time2})). This dual formulation enables observations to simultaneously constrain both the admissible spacetime geometries and their underlying matter content, without specifying a particular metric or matter model or requiring model–by–model analysis. Crucially, these bounds are immediately applicable to current and forthcoming BH observations. 

To illustrate their observational relevance, consider the critical-radial-NEC class, for which the shadow radius satisfies $\eta_{\rm p}=\sqrt{3r_{\rm p}^2/(1-8\pi r_{\rm p}^2\,\rho_{\rm p})}$. Combining this with the 2017 EHT measurement for M87$^{\star}$, $\eta_{\rm p}=3\sqrt{3}M(1+\delta)$ with $\delta=-0.01\pm0.17$ \cite{EventHorizonTelescope:2019ggy}, Eq.~(\ref{Eq:PS_Energy_Density}) yields a joint bound on UPS compactness and local energy density 
$$0.743 \lesssim \left[\left(\frac{3M}{r_{\rm p}}\right)^2-8\pi (3M)^2\rho_{\rm p}\right]\lesssim 1.487.$$ 
Imposing the WEC, $\rho_{\rm p}\geq 0$, further gives the BH model-independent bound on the UPS radius of M87$^{\star}$, $r_{\rm p}\lesssim 3.48M$, while the additional assumption $r_{\rm p}>2M$ constrains the local energy density to $0\leq\rho_{\rm p}\lesssim 0.00666/M^2$. Thus, existing shadow measurements, combined with ECs, already constrain the otherwise inaccessible PS geometry and local matter content, demonstrating the observational potential of EC–driven bounds in the strong–field regime. 

\begin{figure*}[t]
\captionsetup{format=plain, justification=justified}
    \begin{subfigure}{0.49\linewidth}
        \includegraphics[width=\linewidth]{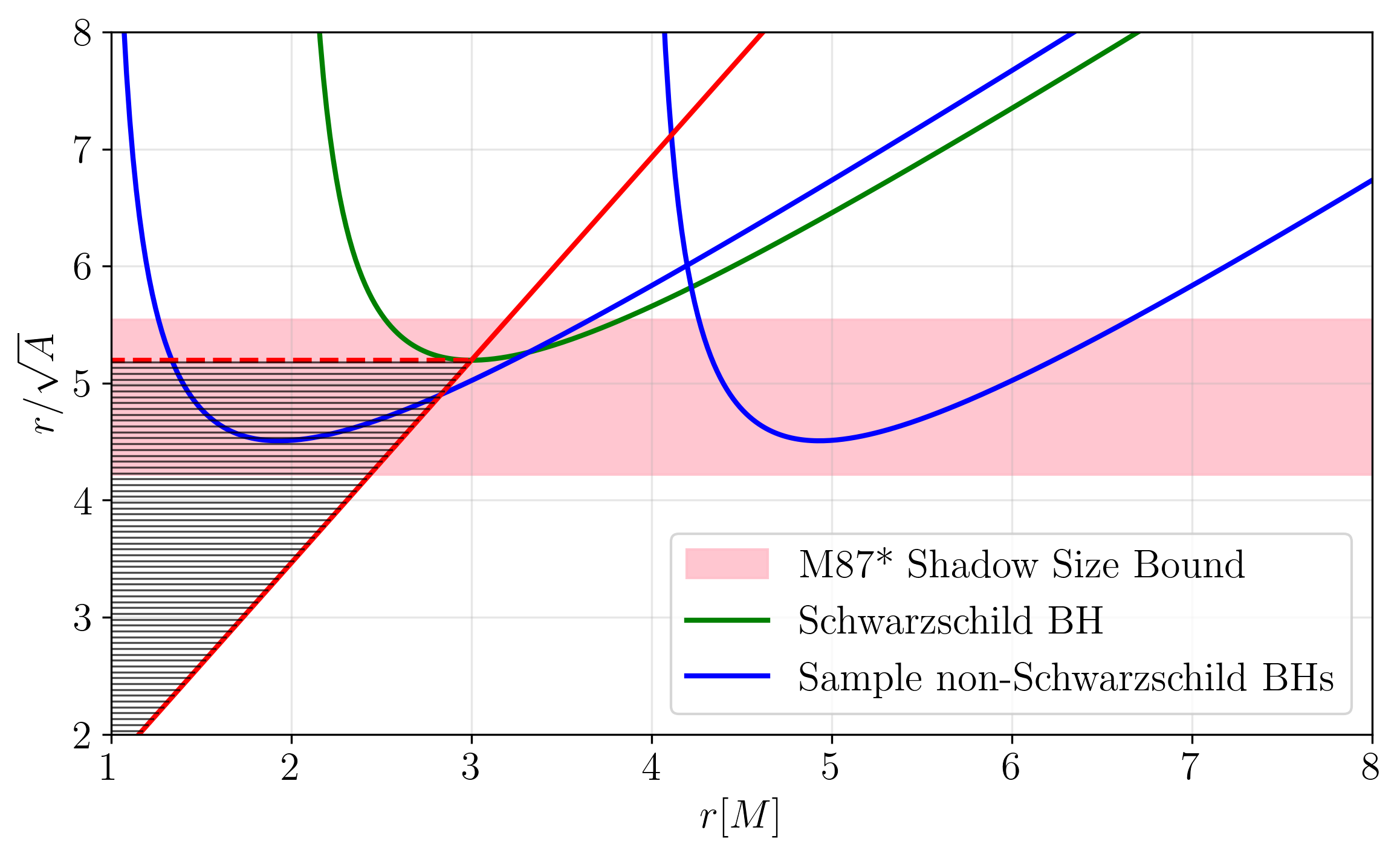}
        \caption{}
        \label{fig:constraints:a}
    \end{subfigure}
    \begin{subfigure}{0.49\linewidth}
        \includegraphics[width=\linewidth]{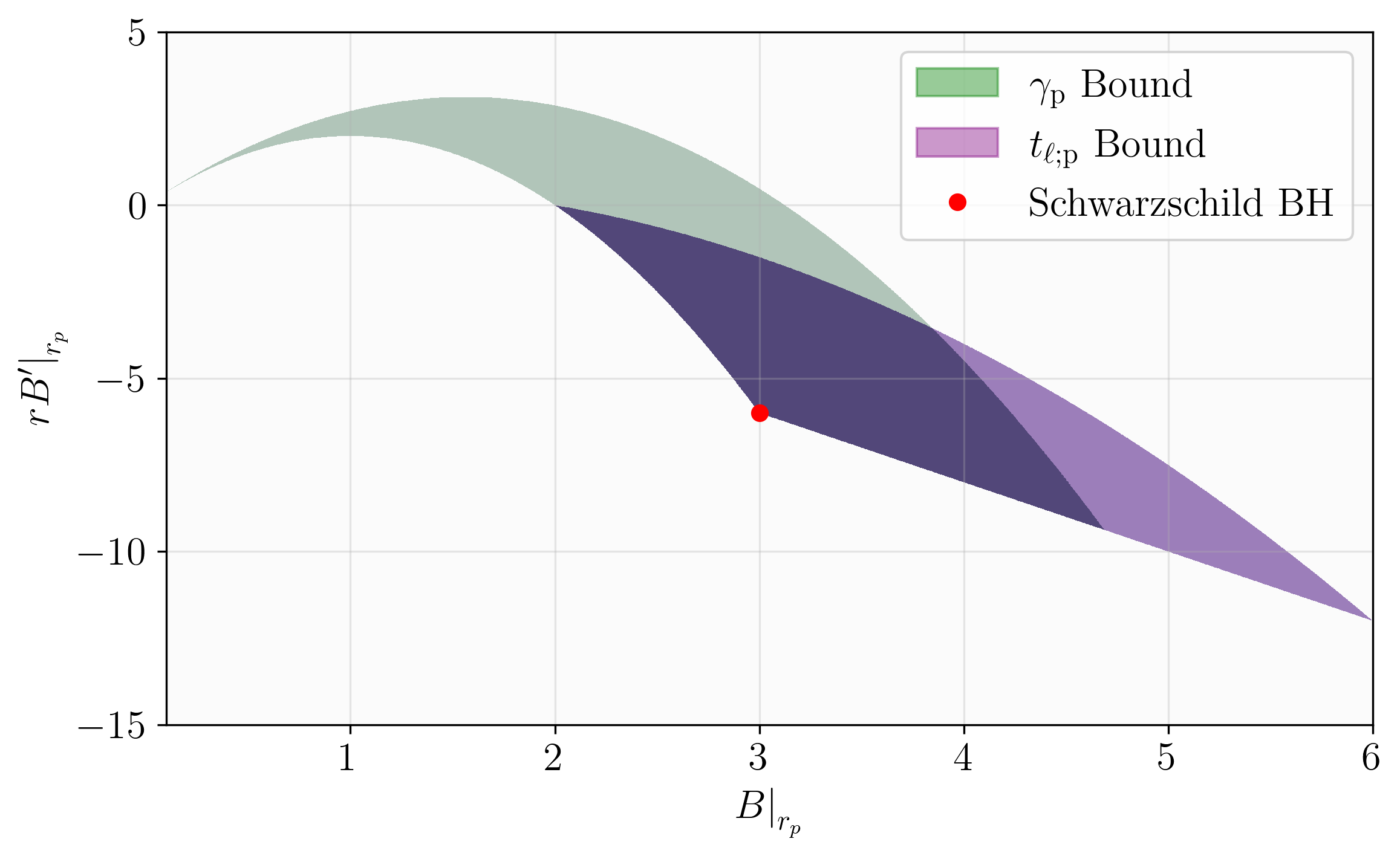}
        \caption{}
        \label{fig:constraints:b}
    \end{subfigure}
    \begin{subfigure}{0.49\linewidth}
        \includegraphics[width=\linewidth]{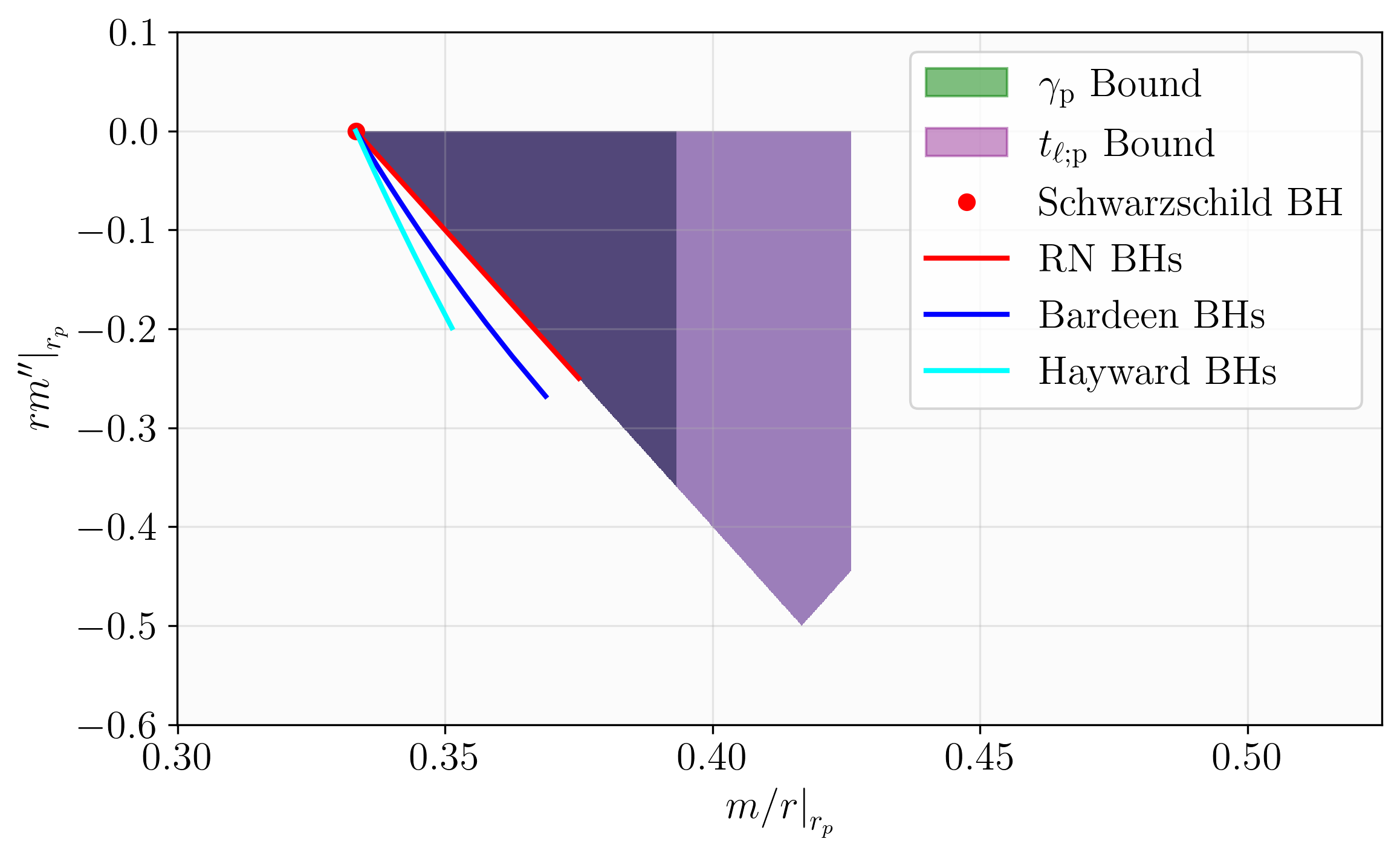}
        \caption{}
        \label{fig:constraints:c}
    \end{subfigure}
    \begin{subfigure}{0.49\linewidth}
        \includegraphics[width=\linewidth]{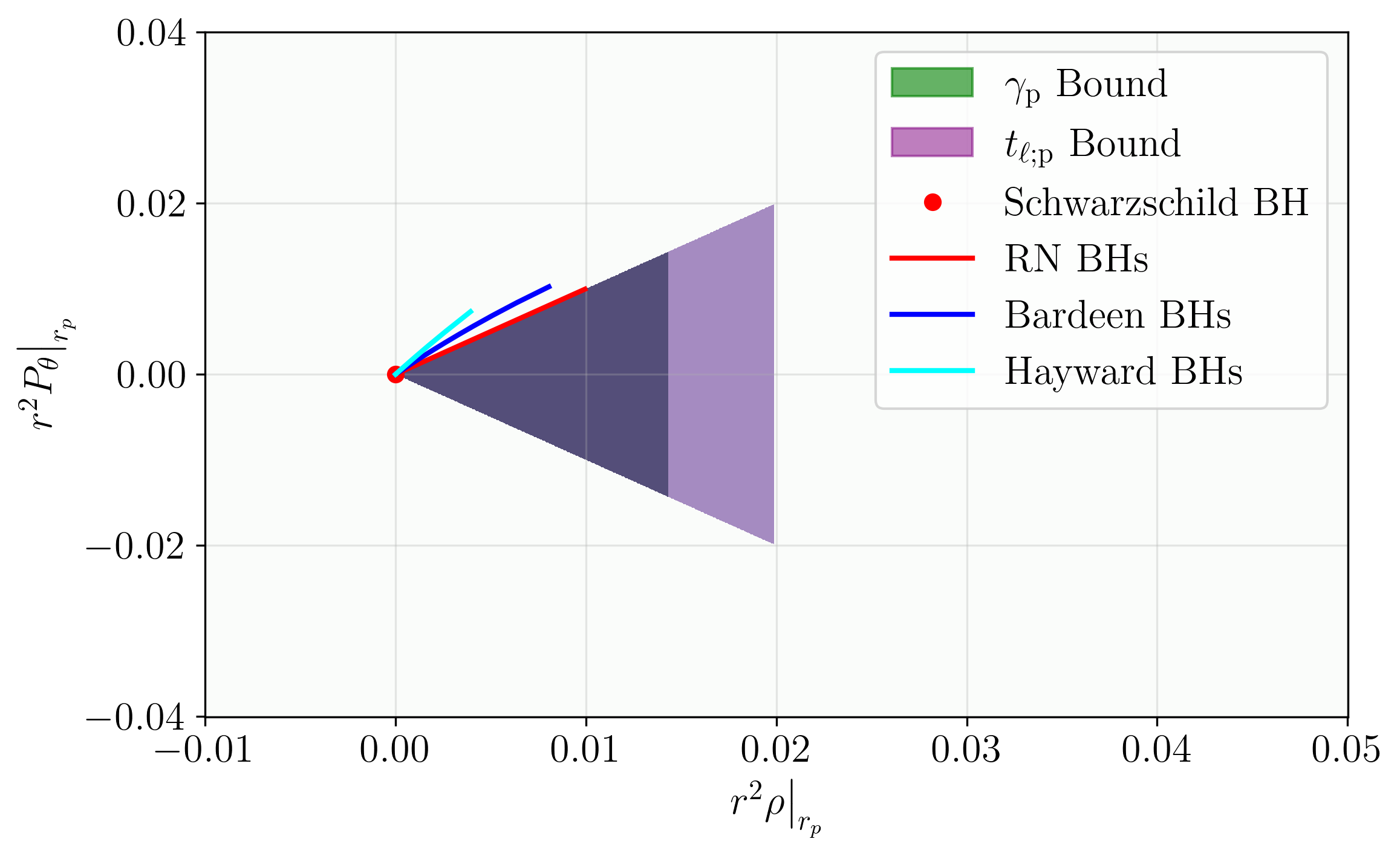}
        \caption{}
        \label{fig:constraints:d}
    \end{subfigure}
    \caption{\justifying
        (a)~Representative impact parameter profiles for static, spherically symmetric BHs. The minimum of each curve determines the UPS radius $r_{\rm p}$ and the shadow radius $\eta_{\rm p}$. An observed shadow size, illustrated by the EHT measurement of M87$^{\star}$ (pink band), is generally consistent with a broad family of BH solutions (colored curves). However, imposing EC-driven bounds (gray shaded region) substantially reduces the allowed parameter space and excludes otherwise viable models.   
        (b)~EC-driven bounds combined with the observationally measured values of the Lyapunov exponent $\gamma_{\rm p}$ and Lyapunov time $t_{\ell; \rm p}$ tightly constrain $B_{\rm p}$ and $r_{\rm p}B_{\rm p}'$. The shaded regions illustrate representative constraints corresponding to exemplary measurements within $20\%$ of the Schwarzschild values. Together they constrain the full spacetime metric. 
        (c)~The resulting observationally inferred constraints on the MSH mass $m$ and its radial derivative for degenerate spacetimes $B(r)=1/A(r)$. 
        (d)~The corresponding constraints on $\rho$ and $P_{(\vartheta)}$ at the PS for degenerate spacetimes. While the Reissner-Nordstrom (RN) BHs satisfy all bounds, the Bardeen and Hayward BHs---owing to their EC violation---fall outside the allowed region. }
    \label{fig:constraints}
\end{figure*}

As shown in Fig.~(\ref{fig:constraints}), current EHT shadow measurements leave substantial degeneracies among BH models, which can be significantly reduced by imposing ECs at the UPS. Future photon ring demagnification measurements can further constrain the admissible BH models and their supporting matter distributions. Using representative EC-satisfying and EC-violating BH solutions, Fig.~(\ref{fig:constraints}) demonstrates the broad applicability and discriminatory power of these bounds. Our framework thus provides a model-independent strategy for constraining both known and yet-to-be-discovered BH and HUCO solutions. Within the class of static, spherically symmetric spacetimes in GR, any observational violations of these bounds would challenge either the assumed ECs or the underlying gravitational and geometric assumptions. The framework can be extended to modified gravity via an effective stress-energy tensor and potentially generalized to rotating, axisymmetric spacetimes, broadening the scope of observational tests of ECs in strong-field gravity.

Although the BH observables $\eta_{\rm p}$, $\gamma_{\rm p}$, $t_{\rm d;p}$, and $t_{\ell;\rm p}$ are geometrically well defined, their precise observational measurement remains challenging. Inferring $\eta_{\rm p}$ requires calibrating the observed emission-ring size to the underlying critical impact parameter (see Sec. 5.3 of \cite{EventHorizonTelescope:2019ggy}), while measuring $\gamma_{\rm p}$ requires resolving successive photon subrings, and extracting $t_{\ell;\rm p}$ from GW ringdown signals requires identifying the relevant QNM modes. However, measuring time delays between successive photon subrings may circumvent the need to resolve their angular separation, while the QNM damping-to-oscillation ratio ($\omega_I/\omega_R$) provides a complementary probe without requiring independent high-precision measurements of both QNM frequency components. Furthermore, forthcoming high-resolution VLBI arrays and next-generation GW observatories are expected to improve the precision and feasibility of measuring these observables, enabling more stringent observational tests of ECs in strong-field gravity.

\begin{acknowledgments}
\emph{Acknowledgments---}The authors thank Prashant Kocherlakota for helpful discussions and the anonymous referee for constructive comments. RKW acknowledges support from the Royal Society Newton International Fellowship (Award No.~NIF\textbackslash R1\textbackslash 251731). Part of RKW's research was conducted at the University of Arizona with support from the Fulbright-Nehru Postdoctoral Research Fellowship (Award No.~2847/FNPDR/2022).
\end{acknowledgments}

\begin{appendix}
\section{Appendix}
\begin{figure*}[h!]
    \centering
    \includegraphics[width=1\linewidth]{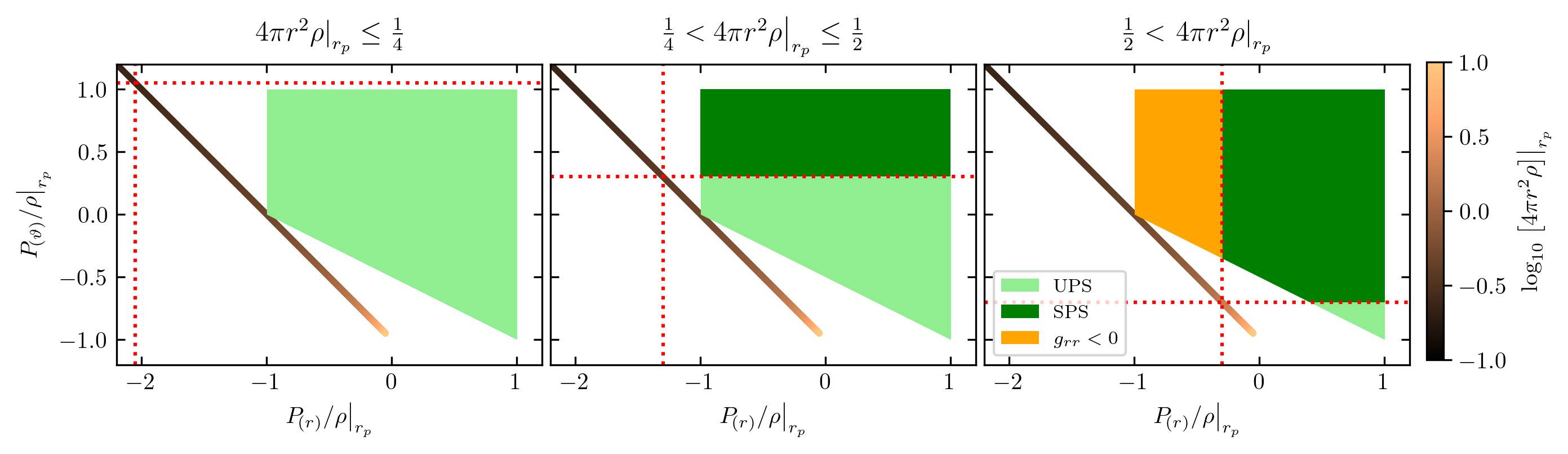}
    \caption{\justifying The shaded region represents EC–satisfying normalized radial and tangential pressure of background gravitating fields at the PS. For critical–radial–NEC spacetimes, admissible configurations lie along $P_{(r)}/\rho=-1$, with $P_{(\vartheta)}/\rho\in(0,1)$. The vertical dashed line separates PSs in static ($g_{rr}>0$, right) and nonstatic ($g_{rr}<0$, left) regions, while the horizontal dashed line distinguishes UPSs (below) from SPSs (above), the latter requiring greater tangential pressure. Varying $\rho_{\rm p}$ traces the solid curve defined by these constraints, which intersects the EC-satisfying region at a unique degenerate point. This construction identifies matter configurations capable of supporting UPSs or SPSs in BH exteriors and HUCO spacetimes.}
    \label{fig:EC2}
\end{figure*}

The classical ECs are used to impose local constraints on the energy density, principal pressures, and metric functions evaluated at the PS. The complete set of inequalities is summarized in Tables~\ref{Table-Static2} and~\ref{Table-Static3}, with a graphical representation shown in Fig.~\ref{fig:EC2}. These constraints for degenerate spacetimes are presented in Table~\ref{Table-Static1}. Beyond these general EC constraints, the (in)stability of the PS yields additional nontrivial restrictions on $\rho$, $\Pi$, and consequently on the metric functions $A$ and $B$. While the NEC and SEC are trivially satisfied at a SPS (cf. Eq.~\eqref{Eq:PS-Stability2}), they need not hold at an UPS. Consequently, a PS that satisfies NEC and SEC may still be stable or unstable. Additionally, $\rho\geq 0$ at the PS implies the universal bound $\rp\leq \left.3m\right|_{\rp}$, regardless of its stability. Among all ECs, the DEC yields the strongest constraints on the MSH mass and the tangential pressure. 

Coordinate-invariant curvature scalars provide complementary probes of the strong-field geometry. Since the Ricci scalar and Ricci square can be written entirely in terms of the local energy density and principal pressures,
\begin{equation}
    \mathscr{R}=8\pi (\rho-P_{(r)}-2P_{\vartheta}),\quad \mathscr{R}^2=64\pi^2 (\rho^2 + P_{(r)}^2 + 2 P_{\vartheta}^2).\nonumber
\end{equation}
At the UPS (SPS), they satisfy
\begin{eqnarray}
    \left.\mathscr{R}\right|_{\rp} & >(<)& \left.\frac{1}{r^3}(6r m' + 6m -4r)\right|_{\rp},\nonumber\\
    \left.\mathscr{R}^2\right|_{\rp} &>(<)& \left.\frac{1}{r^6}(12m(3m-2r)+2r^2(6m'^2-4m'+3))\right|_{\rp},\nonumber
\end{eqnarray}
the EC bounds derived in this work immediately translate into corresponding inequalities on these curvature invariants. These relations furnish model-independent consistency conditions for any static, spherically symmetric BH solution in GR and provide a convenient framework for identifying departures from the GR predictions in the strong-field regime.

\begin{table*}
\centering
\resizebox{\textwidth}{!}{%
\begin{tabular}{|c|c|c|c|}
\hline
        &   \begin{tabular}[c]{@{}c@{}}PS\\  $\mathscr{f}(\rp)=0$ \end{tabular}
        &   \begin{tabular}[c]{@{}c@{}}SPS\\ $\mathscr{f}(\rp)=0$\,\,\,\& \\ $\frac{1}{8\pi \rp^2}<\left.(\rho+\Ptheta)\right|_{\rp}$ \end{tabular}
        &   \begin{tabular}[c]{@{}c@{}}UPS\\ $\mathscr{f}(\rp)=0$\,\,\,\& \\ $\left.(\rho+\Ptheta)\right|_{\rp}<\frac{1}{8\pi \rp^2}$ \end{tabular}
        \\  \hline
NEC     &   $\text{max}\left.\left(-\Ptheta,\, -P_{(r)}\right)\right|_{\rp} \leq \left.\rho\right|_{\rp}$
        &   $\text{max}\left.\left(\frac{1}{8\pi r^2}-\Ptheta,\, -P_{(r)}\right)\right|_{\rp} < \left.\rho\right|_{\rp}$
        &   $\text{max}\left.\left(-\Ptheta,\, -P_{(r)}\right)\right|_{\rp} \leq \left.\rho\right|_{\rp}< \left.\left(\frac{1}{8\pi
            r^2}-\Ptheta\right)\right|_{\rp}$
        \\ \hline
WEC     &   $ \text{max}\left.\left(0,\, -\Ptheta,\, -P_{(r)}\right)\right|_{\rp} \leq \left.\rho\right|_{\rp}$
        &   $\text{max}\left.\left(0,\, \frac{1}{8\pi r^2}-\Ptheta,\, -P_{(r)}\right)\right|_{\rp} < \left.\rho\right|_{\rp}$
        &   $\text{max}\left.\left(0,\, -\Ptheta,\, -P_{(r)}\right)\right|_{\rp} \leq \left.\rho\right|_{\rp} <
            \left.\left(\frac{1}{8\pi r^2}-\Ptheta\right)\right|_{\rp}$
        \\ \hline
SEC     &    $\text{max}\left.\left(-\Ptheta,\, -P_{(r)}, -\Ptheta  -P_{(r)} \right)\right|_{\rp} \leq
            \left.\rho\right|_{\rp}$
        &   $\text{max}\left.\left(-\Ptheta,\, -P_{(r)}, -\Ptheta + \frac{1}{8\pi r^2}\right)\right|_{\rp} \leq
            \left.\rho\right|_{\rp}$
        &   $\text{max}\left.\left(-\Ptheta,\, -P_{(r)}, -\Ptheta  -P_{(r)} \right)\right|_{\rp} \leq
            \left.\rho\right|_{\rp} <  \left.\left(\frac{1}{8\pi r^2}-\Ptheta\right)\right|_{\rp}$
        \\ \hline
DEC     &    $\text{max}\left.\left(0,\, \left|P_{(r)}\right|,\, \left|\Ptheta\right| \right)\right|_{\rp} \leq
            \left.\rho\right|_{\rp}$
        &   $\text{max}\left.\left(0,\, \left|P_{(r)}\right|,\, \left|\Ptheta\right|,\, -\Ptheta + \frac{1}{8\pi r^2} \right)\right|_{\rp} \leq
            \left.\rho\right|_{\rp}$
        &   $\text{max}\left.\left(0,\, \left|P_{(r)}\right|,\, \left|\Ptheta\right| \right)\right|_{\rp} \leq
            \left.\rho\right|_{\rp} <  \left.\left(\frac{1}{8\pi r^2}-\Ptheta\right)\right|_{\rp}$
        \\ \hline
All ECs  &   $\text{max}\left.\left(0,\, \left|P_{(r)}\right|,\, \left|\Ptheta\right|,\, -\Ptheta -P_{(r)}\right)\right|_{\rp} \leq
            \left.\rho\right|_{\rp}$
        &   $\text{max}\left.\left(0,\, \left|P_{(r)}\right|,\, \left|\Ptheta\right|,\, -\Ptheta + \frac{1}{8\pi r^2}\right)\right|_{\rp} \leq
            \left.\rho\right|_{\rp}$
        &   $\text{max}\left.\left(0,\, \left|P_{(r)}\right|,\, \left|\Ptheta\right|,\, -\Ptheta - P_{(r)}  \right)\right|_{\rp} \leq
            \left.\rho\right|_{\rp} < \left.\left(\frac{1}{8\pi r^2}-\Ptheta\right)\right|_{\rp}$
        \\ \hline
\end{tabular}
}\caption{\justifying Constraints on the energy density $\rho$ and the tangential pressure $\Ptheta$ of the background field at the PS of radius $\rp$ in the static, spherically symmetric spacetime with $A \neq 1/B$. }\label{Table-Static2}
\end{table*}
\begin{table*}
\resizebox{\textwidth}{!}{%
\begin{tabular}{|c|c|c|c|}
\hline
        &   \begin{tabular}[c]{@{}c@{}}PS\\  $\mathscr{f}(\rp)=0$  \end{tabular}
        &   \begin{tabular}[c]{@{}c@{}}SPS\\ $\mathscr{f}(\rp)=0$\,\,\,\& \\ $1<\left.\frac{r^2 A''}{2A}\right|_{\rp}$    \end{tabular}
        &   \begin{tabular}[c]{@{}c@{}}UPS\\ $\mathscr{f}(\rp)=0$\,\,\,\& \\ $\left.\frac{r^2 A''}{2A}\right|_{\rp} < 1$    \end{tabular}
            \\  \hline
NEC     &   \begin{tabular}[c]{@{}c@{}} $-2 \leq \left. \frac{rB'}{B}\right|_{\rp}$\;\&\;
            \\ $\left.(1-B)\right|_{\rp} \leq \left. \frac{r^2 A''}{2A}\right|_{\rp}$   \end{tabular}
        &   \begin{tabular}[c]{@{}c@{}} $-2 \leq \left. \frac{rB'}{B}\right|_{\rp}$\;\; \&\;\;
            \\ $1 < \left. \frac{r^2 A''}{2A}\right|_{\rp}$    \end{tabular}
        &   \begin{tabular}[c]{@{}c@{}} $-2 \leq \left. \frac{rB'}{B}\right|_{\rp}$\; \& \;
            \\ $\left. (1-B)\right|_{\rp} \leq \left. \frac{r^2 A''}{2A}\right|_{\rp} < 1$    \end{tabular}
            \\ \hline
WEC     &   \begin{tabular}[c]{@{}c@{}} $ \left.\text{max}(-2, 1-B)\right|_{\rp} \leq \left. \frac{rB'}{B}\right|_{\rp}$\;\&\;
            \\ $\left.(1-B)\right|_{\rp} \leq \left. \frac{r^2 A''}{2A}\right|_{\rp}$   \end{tabular}
        &   \begin{tabular}[c]{@{}c@{}} $\left.\text{max}(-2, 1-B)\right|_{\rp} \leq \left. \frac{rB'}{B}\right|_{\rp}$\;\&\;
            \\ $1 < \left. \frac{r^2 A''}{2A}\right|_{\rp}$   \end{tabular}
        &   \begin{tabular}[c]{@{}c@{}} $\left.\text{max}(-2, 1-B)\right|_{\rp} \leq \left. \frac{rB'}{B}\right|_{\rp}$\;\&\;
            \\  $\left. (1-B)\right|_{\rp} \leq \left. \frac{r^2 A''}{2A}\right|_{\rp} < 1$  \end{tabular}
            \\ \hline
SEC     &   \begin{tabular}[c]{@{}c@{}} $-1 \leq \left. \frac{rB'}{2B}\right|_{\rp}\leq \left.\left(\frac{r^2 A''}{2A}+1\right)\right|_{\rp}$\;\&\;
            \\  $\left.(1-B)\right|_{\rp} \leq \left. \frac{r^2 A''}{2A}\right|_{\rp}$ \end{tabular}
        &   \begin{tabular}[c]{@{}c@{}} $-1\leq \left. \frac{r B'}{2B}\right|_{\rp} \leq \left.\left(\frac{r^2 A''}{2A}+1 \right)\right|_{\rp}$ \;\&\;
            \\  $1 < \left. \frac{r^2 A''}{2 A}\right|_{\rp},$ \end{tabular}
        &   \begin{tabular}[c]{@{}c@{}} $-1\leq \left.\frac{r B'}{2 B}\right|_{\rp}\leq \left.\left(\frac{r^2 A''}{2 A}+1 \right)\right|_{\rp}$\;\&\;
            \\  $\left. (1-B)\right|_{\rp} \leq \left. \frac{r^2 A''}{2 A}\right|_{\rp} < 1$ \end{tabular}
            \\ \hline
DEC     &   \begin{tabular}[c]{@{}c@{}} $\left.\left(1-B+|B-3|\right)\right|_{\rp}\leq \left.\frac{rB'}{B}\right|_{\rp}$,\,\,\&
            \\  $\left.(1-B)\right|_{\rp}\leq \left.\frac{r^2 A''}{2 A}\right|_{\rp} \leq \left.\left(\frac{2rB'}{B}+(B-1)\right)\right|_{\rp}$       \end{tabular}
        &   \begin{tabular}[c]{@{}c@{}} $\left.\left(1-B+|B-3|\right)\right|_{\rp}\leq \left.\frac{rB'}{B}\right|_{\rp}$,\,\,\&
            \\  $1 < \left.\frac{r^2 A''}{2A}\right|_{\rp} \leq \left.\left(\frac{2rB'}{B}+(B-1)\right)\right|_{\rp}$       \end{tabular}
        &   \begin{tabular}[c]{@{}c@{}} $\left.\left(1-B+|B-3|\right)\right|_{\rp}\leq \left.\frac{rB'}{B}\right|_{\rp}$,\,\,\&
            \\  $\left.(1-B)\right|_{\rp}\leq \left.\frac{r^2 A''}{2A}\right|_{\rp} \leq \left.\text{min}\left(1, \frac{2rB'}{B}+(B-1)\right)\right|_{\rp}$        \end{tabular}
            \\ \hline
All ECs  &   \begin{tabular}[c]{@{}c@{}} $\left.\text{max}\left(-2, 1-B+|B-3|\right)\right|_{\rp}\leq \left.\frac{rB'}{B}\right|_{\rp}\leq \left.\left(\frac{r^2A''}                        {A}+2 \right)\right|_{\rp}$\,\,\&
            \\  $\left.(1-B)\right|_{\rp}\leq \left.\frac{r^2 A''}{2 A}\right|_{\rp} \leq \left.\left(\frac{2rB'}{B}+(B-1)\right)\right|_{\rp}. $ \end{tabular}
        &   \begin{tabular}[c]{@{}c@{}} $\left.\text{max}\left(-2, 1-B+|B-3|\right)\right|_{\rp}\leq
            \left.\frac{rB'}{B}\right|_{\rp}\leq \left.\left(\frac{r^2 A''}{A}+2 \right)\right|_{\rp}$\,\,\&
            \\  $1 < \left.\frac{r^2 A''}{2A}\right|_{\rp} \leq \left.\left(\frac{2rB'}{B}+(B-1)\right)\right|_{\rp}$ \end{tabular}
        &   \begin{tabular}[c]{@{}c@{}} $\left.\text{max}\left(-2, 1-B+|B-3|\right)\right|_{\rp}\leq
            \left.\frac{rB'}{B}\right|_{\rp}\leq \left.\left(\frac{r^2A''}{A}+2 \right)\right|_{\rp}$,\,\,\&
            \\  $\text{max}\left.\left(-2, 1-B\right)\right|_{\rp}\leq \left.\frac{r^2 A''}{2A}\right|_{\rp} \leq \left.\text{min}\left(1, \frac{2rB'}{B}+(B-1)\right)\right|_{\rp}$ \end{tabular}
            \\ \hline
\end{tabular}
}\caption{\justifying Constraints on the metric functions, $A$ and $B$, at the PS of radius $\rp$ in the static, spherically symmetric spacetime with $A \neq 1/B$ and $A,B>0$. These constraints are complementary to those presented in Table~ \ref{Table-Static2}.}\label{Table-Static3}
\end{table*}

\begin{table*}
\resizebox{\textwidth}{!}{%
\begin{tabular}{|c|c|c|c|}
\hline
        &   \begin{tabular}[c]{@{}c@{}} PS  \\ $\left.rm'\right|_{\rp}=\left.(3m-r)\right|_{\rp}$ \end{tabular}
        &   \begin{tabular}[c]{@{}c@{}} SPS \\ $\left.rm'\right|_{\rp}=\left.(3m-r)\right|_{\rp}$\,\,\,\& 
            \\ $\left.r^2 m''\right|_{\rp}< \left. -3(r-2m) \right|_{\rp}$ \end{tabular}
        &   \begin{tabular}[c]{@{}c@{}}UPS\\ $\left.rm'\right|_{\rp}=\left.(3m-r)\right|_{\rp}$ \,\,\,\&
            \\ $\left.-3(r-2m) \right|_{\rp} < \left. r^2 m'' \right|_{\rp}$ \end{tabular}
            \\  \hline
NEC     &   $ \left. r^2 m'' \right|_{\rp}\leq \left. -2(r-3m) \right|_{\rp}$
        &   Always
        &   $ \left.-3(r-2m)\right|_{\rp} < \left. r^2 m'' \right|_{\rp} \leq \left. -2(r-3m) \right|_{\rp}$ 
            \\  \hline
WEC     &   \begin{tabular}[c]{@{}c@{}}$\left. \frac{1}{3}\leq \frac{m}{r} \right|_{\rp}$\,\,\, \&\, \\ $ \left. r^2 m'' \right|_{\rp}\leq \left. -2(r-3m) \right|_{\rp}$\end{tabular}
        &   \begin{tabular}[c]{@{}c@{}}$\left. \frac{1}{3}\leq \frac{m}{r} \right|_{\rp}$\,\,\, \&\, \\ $ \left. r^2 m'' \right|_{\rp} < \left. -3(r-2m) \right|_{\rp} $ \end{tabular}
        &   \begin{tabular}[c]{@{}c@{}}$\left. \frac{1}{3}\leq \frac{m}{r} \right|_{\rp}$\,\,\, \&\, \\ $ \left. -3(r-2m)\right|_{\rp} < \left. r^2 m'' \right|_{\rp} \leq \left. -2(r-3m) \right|_{\rp}$\end{tabular}
            \\  \hline
SEC     &   $ \left. r^2 m'' \right|_{\rp} \leq \text{min}\left.\Big(0, -2(r-3m)\Big)\right|_{\rp}$
        &   Always
        &   $ \left. -3(r-2m) \right|_{\rp} < \left. r^2 m'' \right|_{\rp} \leq  \left. \text{min}\Big(0, -2(r-3m)\Big) \right|_{\rp}$    
            \\  \hline
DEC     &   \begin{tabular}[c]{@{}c@{}}$\left. \frac{1}{3}\leq \frac{m}{r} \right|_{\rp}$\,\,\,\&  \\
        $   \left. 2(r-3m) \right|_{\rp} \leq \left. r^2 m'' \right|_{\rp} \leq  \left. -2(r-3m) \right|_{\rp}$  \end{tabular}
        &   \begin{tabular}[c]{@{}c@{}}$\left. \frac{5}{12} < \frac{m}{r} \right|_{\rp}$\,\,\,\, \&\,  \\ $ \left. 2(r-3m) \right|_{\rp} \leq \left. r^2 m'' \right|_{\rp} <  \left. -3(r-2m) \right|_{\rp}$\end{tabular}
        &   \begin{tabular}[c]{@{}c@{}}$\left. \frac{1}{3}\leq \frac{m}{r} \right|_{\rp}$\,\,\,\,\, \&\, \\ $ \left.\text{max}\Big(2(r-3m), -3(r-2m) \Big)\right|_{\rp} < \left.  r^2 m'' \right|_{\rp} \leq \left. -2(r-3m) \right|_{\rp} $\end{tabular}       
            \\  \hline
All ECs  &   \begin{tabular}[c]{@{}c@{}}$\left.\frac{1}{3}\leq \frac{m}{r} \right|_{\rp} < \frac{1}{2}$\,\,\,\, \&\, \\ $ \left. 2(r-3m) \right|_{\rp} \leq \left.  r^2 m'' \right|_{\rp} \leq 0 $ \end{tabular}
        &   \begin{tabular}[c]{@{}c@{}}$\frac{5}{12}<\left. \frac{m}{r} \right|_{\rp}<\frac{1}{2}$\, \&\, \\ $ \left. 2(r-3m)\right|_{\rp} \leq \left. r^2 m'' \right|_{\rp} <  \left. -3(r-2m) \right|_{\rp}$ \end{tabular}
        &  \begin{tabular}[c]{@{}c@{}}$\frac{1}{3}\leq \left. \frac{m}{r} \right|_{\rp}<\frac{1}{2}$\,\,\,\, \&\, \\ $ \left.\text{max}\Big(2(r-3m),\, -3(r-2m)\Big) \right|_{\rp}  <  \left. r^2 m'' \right|_{\rp} \leq 0
            $ \end{tabular}   
            \\  \hline
\end{tabular}}
\caption{\justifying Constraints on the MSH mass function and its derivative at the PS of radius $\rp$ in the static, spherically symmetric spacetime with $A=1/B=1-2m/r$ and $A,B>0$. Stable PSs are generally more compact than unstable ones.}\label{Table-Static1}
\end{table*}

\end{appendix}
\bibliography{Refs}
\end{document}